# New Samarium and Neodymium based admixed ferromagnets with near-zero net magnetization and tunable exchange bias field

Prasanna D. Kulkarni[1], U. V. Vaidya[1], S. K. Dhar[1], P. Manfrinetti[2] and A. K. Grover[1]

[1]Department of Condensed Matter Physics and Materials Science, Tata Institute of Fundamental Research, Colaba, Mumbai-400005.
[2]INFM and Dipartimento di Chimica e Chimica Industriale, Università di Genova, Via Dedecaneso 31, Genova 16146, Italy

e-mail: prasanna@tifr.res.in

## Abstract

Rare earth based intermetallics, SmScGe and NdScGe, are shown to exhibit near zero net magnetization with substitutions of 6 to 9 atomic % of Nd and 25 atomic % of Gd, respectively. The notion of magnetic compensation in them is also elucidated by the crossover of 'zero magnetization' axis at low magnetic fields (< $10^3$ Oe) and field-induced reversal in the orientation of the magnetic moments of the dissimilar rare earth ions at higher magnetic fields. These magnetically ordered materials with 'no-net' magnetization and appreciable conduction electron polarization display an attribute of an exchange bias field, which can be tuned. The attractively high magnetic ordering temperatures of ~ 270 K, underscore the importance of these materials for potential applications in spintronics.

## 1. Introduction

The interest in tailoring Samarium based ferromagnets, having near-zero net magnetization along with large conduction electron polarization, for applications in spintronics has gained momentum with the realization of $Sm_{1-x}Gd_xAl_2$ alloys ($0.01 \leq x \leq 0.032$), in the thin film form [1,2]. These admixed alloys undergo magnetic ordering at ~ 125 K. Adachi and Ino [3] had pointed out that amongst rare earth based systems, the Samarium alloys are particularly suitable for conceiving materials having seemingly incompatible properties of large spin polarization and no net magnetization. From application point of view, it would be fruitful to search for other ferromagnetic Samarium and rare earth ($R$) based intermetallics, which have magnetic ordering temperatures above or close to the ambient temperature and drive them towards the near zero magnetization limit by suitable admixing. In the present work, we identify two such magnetically ordered systems, which sustain near-zero magnetization values up to temperatures much higher than 200 K. Typically, the zero magnetization state amongst the metallic systems is achieved in admixed rare earth alloys with the two R ions belonging to the two different halves of the 4$f$-rare earth series (such that the $S$-state ($L = 0$) $Gd^{3+}$ ions ($S = 7/2$) are counted along with the $R^{3+}$ ions belonging to second half of the 4$f$-series). In such admixed systems, the magnetic moments of dissimilar R ions belonging to the different halves of the 4$f^{\,n}$ series (for which $J = L-S/L+S$ in the ground multiplet of the first half / second half) get antiferromagnetically linked via the indirect exchange interaction mediated by the conduction electrons. In an isoelectronic $R$-series of alloys, which display ferromagnetic ordering, the indirect exchange interaction maintains ferromagnetic coupling between the "spins" of the 4$f$-series of $R^{3+}$ ions under all circumstances [4,5]. However, the magnetizations of the "magnetic moments" of dissimilar $R$ ions occupying the same crystallographic site in a given matrix evolve differently with temperature, and specific stoichiometries in the admixed rare earth alloys are well known [4,6] to yield magnetic compensation behaviour. In recent years [3,7], it has been anticipated that the magnetization ($M$) response per Samarium ion, aided by a contribution from conduction electron polarization, in a given Samarium intermetallic can either be orbital-surplus or spin-surplus and it can be induced to the zero magnetization limit either on substitution of $Sm^{3+}$ ions by $Gd^{3+}$ or $Nd^{3+}$ ions, as already reported in $SmAl_2$ [3,9-13] and SmCd systems [7,14], respectively.

Several intermetallic compounds in the two ferromagnetic rare earth ($R$) series, RScGe and RScSi, have ordering temperatures well above 200 K [15]. We are able to tune the magnetization of two compounds of SmScGe and NdScGe across the $M = 0$ axis by 9 at. % substitution of $Sm^{3+}$ by $Nd^{3+}$ ions and by 25 at. % replacement of $Nd^{3+}$ by $Gd^{3+}$, respectively. The characteristic results of magnetization measurements in $Sm_{1-x}Nd_xScGe$ ($x = 0.06$ and 0.09) and $Nd_{0.75}Gd_{0.25}ScGe$, having $T_c$ ~ 270 K and ~ 240 K, respectively have been chosen for presentation here. The $M$ values in $Sm_{0.94}Nd_{0.06}ScGe$ remain very close to zero even in very large field (50 kOe) and this composition would be particularly suitable for applications, where self-stray field is of concern. We are also reporting here on the attribute of exchange bias fields in these alloys on the basis of detailed magnetization hysteresis studies in them. The exchange bias, which can be tuned in the present series of alloys, enhance the potential use of these materials in magnetic multi-layer composites for specific device applications.

## 2. Experimental

Polycrystalline samples of $Sm_{1-x}Nd_xScGe$ ($x = 0.06, 0.09$) were prepared by melting together the stoichiometric amounts of SmScGe and NdScGe. In the case of $Nd_{1-x}Gd_xScGe$ ($x = 0.25$), the stoichiometric amounts of the pure elements were melted together in an arc furnace. The characterization of the powdered samples by x-ray diffraction confirmed their CeScGe type tetragonal structure. The magnetization data were obtained using Quantum Design Inc. SQUID Magnetometers (Model MPMS 5 and SVSM).

## 3. Results and Discussions

Figures 1 (a) and 1 (b) show the field cooled cool-down (FCC) magnetization curves ($M_{FCC}$) in $Sm_{0.94}Nd_{0.06}ScGe$ and $Sm_{0.91}Nd_{0.09}ScGe$ at low and high applied fields, respectively. The shapes of the $M_{FCC}(T)$ curves in Figs. 1(a) and 1(b) display a non-monotonic behaviour, and the low values (< 0.15 $\mu_B$/formula unit) imply the existence of a competition between magnetic responses from different possible constituents contributing to the total *M* signal. The magnetic moments per formula unit (f.u.) in the ordered state in the pristine ferromagnets $R$ScGe ($R$ = Sm, Nd and Gd) had been reported to be ~ 0.2, 2 and 7 $\mu_B$ [15].

The onset of magnetic ordering near 270 K associated with the host Sm-matrix in the two stoichiometries ($x$ = 0.06, 0.09) can be easily identified from their low field $M_{FCC}(T)$ curves in Fig. 1(a). Both the curves in Fig. 1(a) display a similar peak like response below $T_c$, however, there is one significant difference. The $M_{FCC}(T)$ curve for $x$ = 0.09 can be seen to crossover to the metastable negative values below compensation temperature ($T_{comp}$) of about 88 K, whereas in the $M_{FCC}(T)$ curve for $x$ = 0.06, the values reach a minimum limit at about 50 K, and thereafter the curve flattens out. The $M_{FCC}(T)$ curve in 50 kOe in $x$ = 0.06 alloy in Fig. 1(b) shows a turnaround in magnetization response at about 90 K (identified as $T^*$), thereby elucidating the compensation behaviour between the two competing contributions at this composition as well. The $M_{FCC}(T)$ curve for $x$ = 0.09 in $H$ = 50 kOe shows a shallower minimum and a change in curvature at about 180 K (being marked as $T^*$ in Fig. 1(b)). *M* values in $x$ = 0.06 alloy just vary from ~ 0.015 $\mu_B$ / f.u. to ~ 0.05 $\mu_B$ / f.u. from 300 K down to zero temperature, this attests to its enhanced potential for applications, when stray field due to sample magnetization is of particular concern.

Fig. 2 shows the $M_{FCC}(T)$ curves in 50 Oe and 50 kOe in Nd$_{0.75}$Gd$_{0.25}$ScGe. The low field $M_{FCC}(T)$ curve displays a $T_c$ near 240 K and zero crossover at $T_{comp}$ ~ 105 K, whereas in a higher field of 50 kOe, the turnaround behaviour is witnessed at $T^*$ of ~ 120 K. Both the characteristics are akin to those observed in the Sm$_{1-x}$Nd$_x$ScGe series alloys described above. However, the *M* values in 50 kOe in Fig. 2 are somewhat larger as compared to those in Fig. 1(b).

A subtle difference in the magnetic compensation behaviour in Nd$_{0.75}$Gd$_{0.25}$ScGe and Sm$_{0.91}$Nd$_{0.09}$ScGe can be further highlighted by comparing the warm up of the remanent magnetization ($M_{Rem}$) in them (see Figs. 3(a) and 3(b)). $M_{Rem}$ is initially obtained by cooling the sample to 5 K in a field of 50 kOe and then reducing the field to 50 Oe. Note that in Fig. 3(a), $M_{Rem}(T)$ and $M_{FCC}(T)$ curves in the Nd$_{0.75}$Gd$_{0.25}$ScGe compound have mirror images like appearances from 5 K upto ~ $T_c$, the two curves intersect at $T \approx 120$ K (close to the $T^*$ value in Fig. 2). On the other hand, in Fig. 3(b), for Sm$_{0.91}$Nd$_{0.09}$ScGe, *prima facie*, there does not appear any correspondence between the $M_{Rem}(T)$ and $M_{FCC}(T)$ curves. The two curves do, however, intersect at $T \approx 188$ K, and above this temperature $M_{Rem}(T)$ curve lies below $M_{FCC}(T)$ curve. The two curves eventually merge into each other as $T \rightarrow T_c$ ($\approx 270$ K). In fact, a somewhat similar feature of merger into each other while approaching $T_c$ value can be noted for $M_{Rem}(T)$ and $M_{FCC}(T)$ curves in Fig. 3(a) as well. The observations in Fig. 3 can be rationalized by stating that Nd moments in the remanent state at 5 K are oriented along the field direction and such an orientation remains sustained nearly upto the respective $T_c$ values. The magnetization values at the temperatures of intersection of the $M_{Rem}(T)$ and $M_{FCC}(T)$ curves in Fig. 3(a) and Fig. 3(b) are close to the respective magnetization values at the temperatures of merger of two curves near (respective) $T_c$ values. $M_{Rem}(T)$ values between the temperature of intersection and temperature of merger of two curves in Fig. 3(a) and Fig. 3(b) are in a metastable region in the sense that they are lower than the corresponding $M_{FCC}$ values. Similarly, the $M_{FCC}(T)$ values below intersection temperature are also in the metastable region. The collection of higher of the set of two *M* values at a given temperature would constitute a *quasi*-equilibrium *M-T* curve in a field of 50 Oe. Such a collection would display a minimum at a temperature, close to its $T^*$ value, determined from high field.

The net magnetization has contributions from three constituents, viz., the magnetic moments of two rare earth ions and the conduction electron polarization (CEP). The CEP contribution (exchanged coupled to the rare earth spins) is expected to undergo a phase reversal as the rare earth moments undergo a field induced reversal across a given $T^*$. The CEP contribution is expected to be of the order of 0.1 $\mu_B$ / f.u. in both the compounds, however, the net offset values from the two local moments in Sm$_{0.91}$Nd$_{0.09}$ScGe would be much lower than those in the Nd$_{0.75}$Gd$_{0.25}$ScGe, as the former involves competition between smaller moment values (i.e., contribution from 91% of Sm moments and 9% of Nd moments) as compared to those in the latter admixed compound. This probably accounts for the larger difference between the values of $T^*$ and $T_{comp}$ in the Sm based compound as compared to that in the Nd based compound. At $T_{comp}$, the net magnetization from all the three contributions equals zero (at low fields), whereas, $T^*$ perhaps represents a limit, where the contributions (at high fields) from the two local moments balance out prior to reaching the $T_{comp}$ value.

In the rare earth based ferromagnets with near zero net magnetization, one can imagine that the CEP part is akin to a soft component which is exchange coupled to a pseudo-antiferromagnet made up of dissimilar rare earth moments. In Sm$_{0.98}$Gd$_{0.02}$Al$_2$, Chen *et al.* [16] discovered the presence of the exchange bias field ($H_{exch}$ = -($H_+$ + $H_-$)/2), as characterized by the asymmetric shift in the magnetization hysteresis loop below its $T_{comp}$, where $H_+$ and $H_-$ represent the fields at which the *M* crosses the zero value in a *M-H* loop. The temperature variations of $H_{exch}$ and the effective coercive field (defined as half width of the hysteresis loop, $H_c^{eff}$ = ($H_+$ - $H_-$

)/2) in $Sm_{0.98}Gd_{0.02}Al_2$ imbibes a complex non-monotonic behavior. While $H_c^{eff}(T)$ appeared to display a divergence as $T \to T_{comp}$ from low/high temperature ends, the $H_{exch}(T)$ displays a peak, followed by an increase, as temperature was lowered below $T_{comp}$ (see Fig.2 in Ref. 16). We show in Fig. 4 the *M-H* loops in $Sm_{0.94}Nd_{0.06}ScGe$ alloy recorded between ± 70 kOe alloy at three representative temperatures, viz., 5 K, 50 K and 80 K below its nominal $T^*$ of 90 K (cf. *M(T)* in 50 kOe in Fig. 1(b)). The shifts in the centre of gravity of the hysteresis loop in the horizontal as well as vertical directions are well apparent. Such shifts have been reported in the magnetic multi-layer composites comprising ferromagnetic and antiferromagnetic layers as well as in other admixed magnetic systems [17,18]. The insets (a) and (b) in Fig. 4 show the temperature variations of $H_{exch}$ and $H_c^{eff}$ in $Sm_{0.94}Nd_{0.06}ScGe$. While the shape of $H_{exch}(T)$ curve in the inset (a) bears resemblance to its shape in $Sm_{0.98}Gd_{0.02}Al_2$ [16], the $H_c^{eff}(T)$ has a perplexing non-monotonic response, comprising two peaks centered around 100 K and 20 K, respectively, and with a minimum in between at ~ 70 K. The values of the exchange bias field in the inset (a) of Fig. 4 are an order magnitude larger than the corresponding values reported in the $Sm_{0.98}Gd_{0.02}Al_2$ alloy [16]. The effective coercive field values, likewise, are also significantly larger in the alloy under study. The composition $Sm_{0.94}Nd_{0.06}ScGe$ thus emerges as an example of a magnetically ordered material having (i) 'no-net' magnetization, (ii) large spin polarization, (iii) high exchange bias field and (iv) moderate to large effective coercive field.

The change of stoichiometry from $Sm_{0.94}Nd_{0.06}ScGe$ to $Sm_{0.91}Gd_{0.09}ScGe$ allows the possibility of exploring smaller values of the exchange bias field. Fig. 5 shows the magnetization hysteresis loops recorded between ± 10 kOe at few selected temperatures in $Sm_{0.91}Gd_{0.09}ScGe$. The asymmetry in the shape of these loops is well evident at $T \leq 160$ K, it is seen to rapidly enhance as the sample is cooled down below 50 K. It can be noted that while the magnetization values in this sample are higher than the earlier one, the exchange bias fields (i.e., the asymmetric shifts) are significantly lower. The insets (a) and (b) in Fig. 5 show the plots of $H_{exch}(T)$ and $H_c^{eff}(T)$, respectively. A comparison of the inset (b) in Fig. 4 with that in Fig. 5 shows how the effective coercive field values differ in the two compositions. The shapes of $H_{exch}(T)$ curves in the insets (a) of Fig. 4 and Fig. 5 are not too different. It is apparent that the composition variation in the $Sm_{1-x}Nd_xScGe$ (0.06 ≤ *x* ≤ 0.09) series would permit a wide range of selection of the values of exchange field and effective coercive field in ferromagnets with 'no net' magnetization and large spin polarization.

To gain some comprehension of the behavior presented in the insets (b) of Fig. 4 and Fig. 5, we recall the behavior of effective coercive field reported in the literature [19,20] in ferromagnetic materials comprising 4*f*-rare earth and 3*d*-transition metals in which magnetic moments of the rare earth and transition elements couple antiferromagnetically. In particular, Webb *et al.* [20] had found that multilayer composites of Gd and Co prepared in two different ways yielded nearly the same compensation temperature but displayed different temperature dependences for the effective coercive field. The effective coercive field ($H_c^{eff}$) is a material attribute, which is known to depend in a complex way on the details of the composition modulations and multilayering in the Gd-Co composites [19]. In ideal ferrimagnets, the coercive field ($H_c$) is considered to inversely relate to the (extrapolated) spontaneous magnetization ($M_s$) and is expected to diverge as $T \to T_{comp}$ [20,21] to conform with the collapse in $M_s$ at $T_{comp}$, where the contributions from the two different sublattices balance out and the *M* versus *H* plot becomes linear, like, in antiferromagnets. Webb *et al.* [20] found that in the case of (uniform) amorphous phase of Gd-Co composite, the $H_c$ showed a divergence, whereas in the case of (an inhomogeneous) layered Gd-Co composite, $H_c^{eff}(T)$ displayed a minimum value at $T_{comp}$. It was argued [20] that *M-H* loop in the latter case comprised two parts, one of which switched direction like a soft magnetization contribution having unidirectional anisotropy and the other gave a linear *M-H* response. Such a superposition can give effective coercive field to be lower than the 'true' coercive field, where the 'spontaneous' magnetization switches direction. Webb *et al.* [20] provided a phenomenological description to encompass the observations ranging from divergence to minimum in $H_c^{eff}(T)$ in a variety of ferrimagnets in terms of the strength of (antiferromagnetic) coupling between the two compensating components of a ferrimagnet. In such a description, it is conceivable that at an intermediate strength (in between the strong to weak coupling), the effective coercive field can yield complex behavior, comprising a divergence as well as a minimum in the $H_{eff}(T)$ curve. In the admixed rare earth alloys displaying compensation temperature, the two dissimilar rare earth elements occupying the same crystallographic site are, *prime facie*, strongly coupled, however, they do indeed respond as if the molecular field of one does not strongly influence the molecular field of the other and in this sense they should be considered as weakly coupled. In the case of $Sm_{0.94}Nd_{0.06}ScGe$, the minimum in the $H_c^{eff}(T)$ is close to its $T^*$ value of ~ 70 K (cf. inset (b), Fig. 4). The temperature dependence of the (extrapolated) spontaneous magnetization (data not shown here) in this sample yielded a shallow minimum at about 70 K, which is satisfactory (cf. Fig. 5, Ref 20 in Gd-Co composite as well).

## 4. Conclusions

To conclude, we have illustrated the magnetic behaviour of the rare earth based zero magnetization "spin ferromagnets" in $Sm_{1-x}Nd_xScGe$ ($x$ = 0.06 and 0.09) and $Nd_{1-x}Gd_xScGe$ ($x$ = 0.25) having ordering temperatures close to the ambient temperature. The fact that magnetic compensation in SmScGe happens on substitution of $Sm^{3+}$ ions by $Nd^{3+}$ ions, instead of that by $Gd^{3+}$ ions, confirms the spin-surplus character of the (Sm) magnetization in it. The existence of appreciable conduction electron polarization and a tunable exchange bias field concomitant with the negligible stray field due to sample magnetization in the doped SmScGe and NdScGe compounds are attractive attributes for applications in spintronics.

**Figure captions**

Fig. 1. Temperature variations of magnetization ($M$) values in $Sm_{1-x}Nd_xScGe$ ($x$ = 0.06 and 0.09) during field cool cooldown (FCC) process in (a) 50 Oe and (b) 50 kOe. $T_{comp}$ of 88 K in panel (a) identifies the temperature of crossover of $M$ = 0 axis towards negative metastable values in $x$ = 0.09 alloy. $T^*$ in panel (b) identifies the turnaround in magnetization values at high fields in $x$ = 0.06 alloy.
Fig. 2. $M_{FCC}$ curves in $H$ = 50 Oe and 50 kOe in $Nd_{0.75}Gd_{0.25}ScGe$ alloy. $T_c$, $T_{comp}$ and $T^*$ values have been marked.
Fig. 3. Plots of warm up of remanent magnetization ($M_{Rem}$) from 5 K to 300 K alongwith the $M_{FCC}(T)$ curve in a field of 50 Oe in (a) $Nd_{0.75}Gd_{0.25}ScGe$ and (b) $Sm_{0.91}Nd_{0.09}ScGe$. The temperatures of intersection of the two curves in panels (a) and (b) have been identified, and the $T_{comp}$ values have been marked as well.

Fig. 4. Magnetization hysteresis loops in $Sm_{0.94}Nd_{0.06}ScGe$ at three temperatures, as indicated. Samples were cooled to a given temperature in a field of 70 kOe before recording the loop. The insets (a) and (b) show temperature variations of $H_{exch}$ (= $-(H_+ + H_-)/2$) and $H_c^{eff}$ (= $(H_+ - H_-)/2$). A representative pair of $H_+$ and $H_-$ values are marked for the *M-H* curve at 5 K.

Fig. 5. Magnetization hysteresis loops in $Sm_{0.91}Nd_{0.09}ScGe$ at 80 and 160 K, recorded after an initial cooling in 10 kOe to a given temperature. The insets (a) and (b) show temperature variation of $H_{exch}$ and $H_c^{eff}$. $T^*$ ($\approx$ 180 K) has also been marked in the both insets.

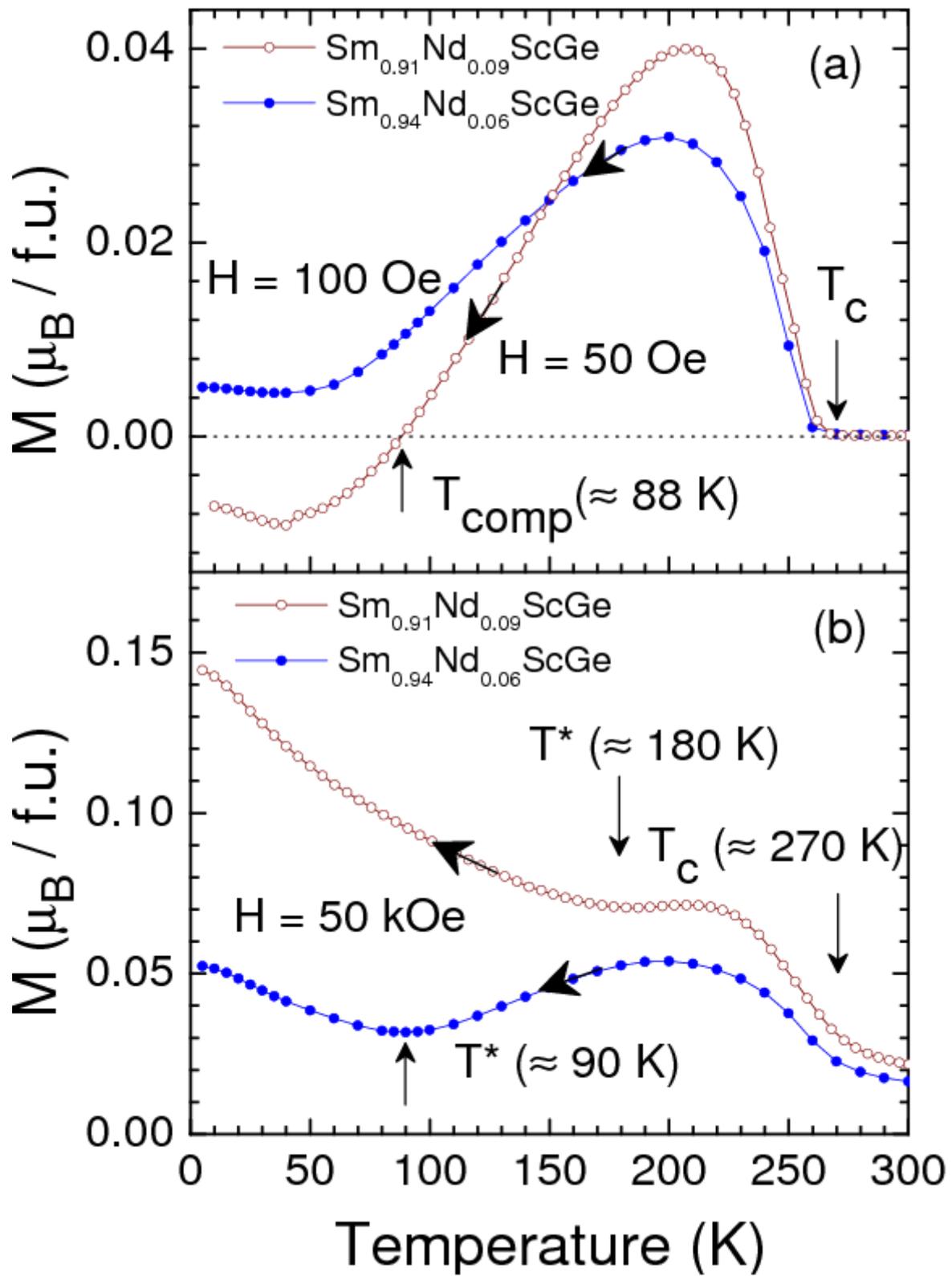

Fig. 1.

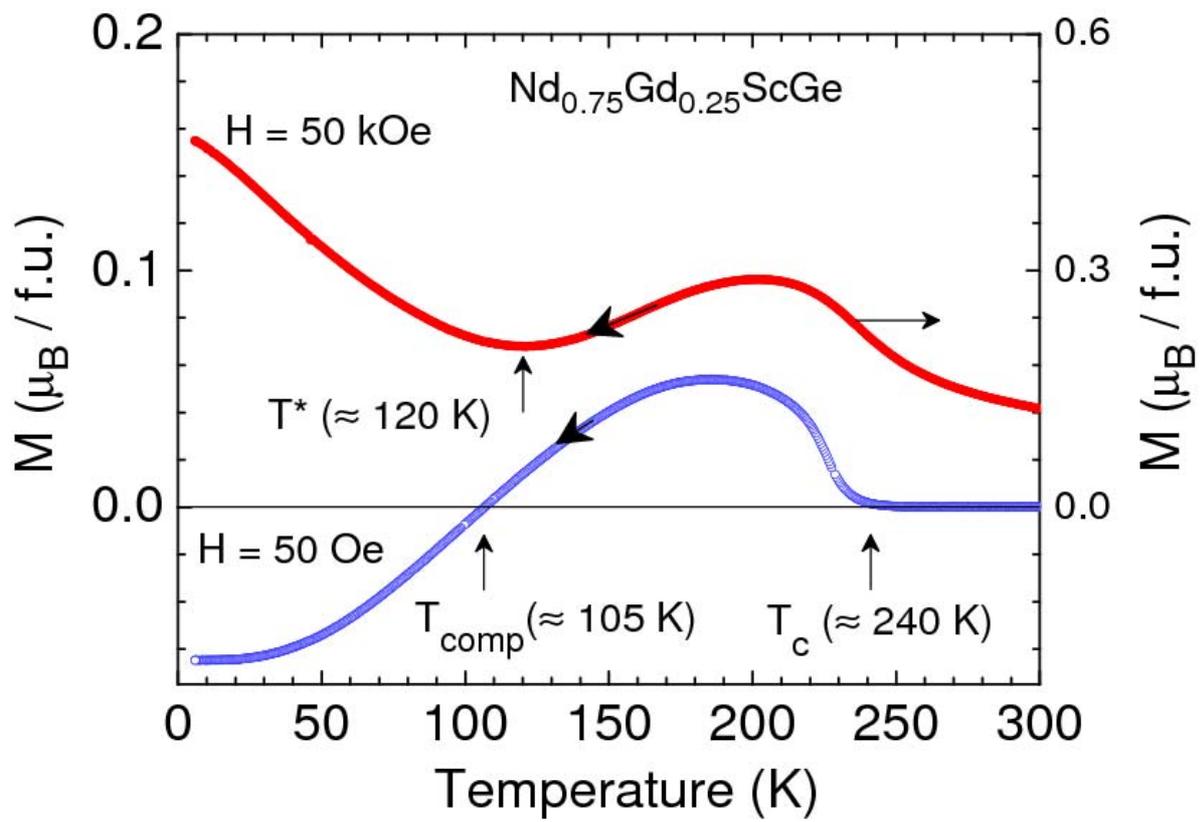

Fig. 2.

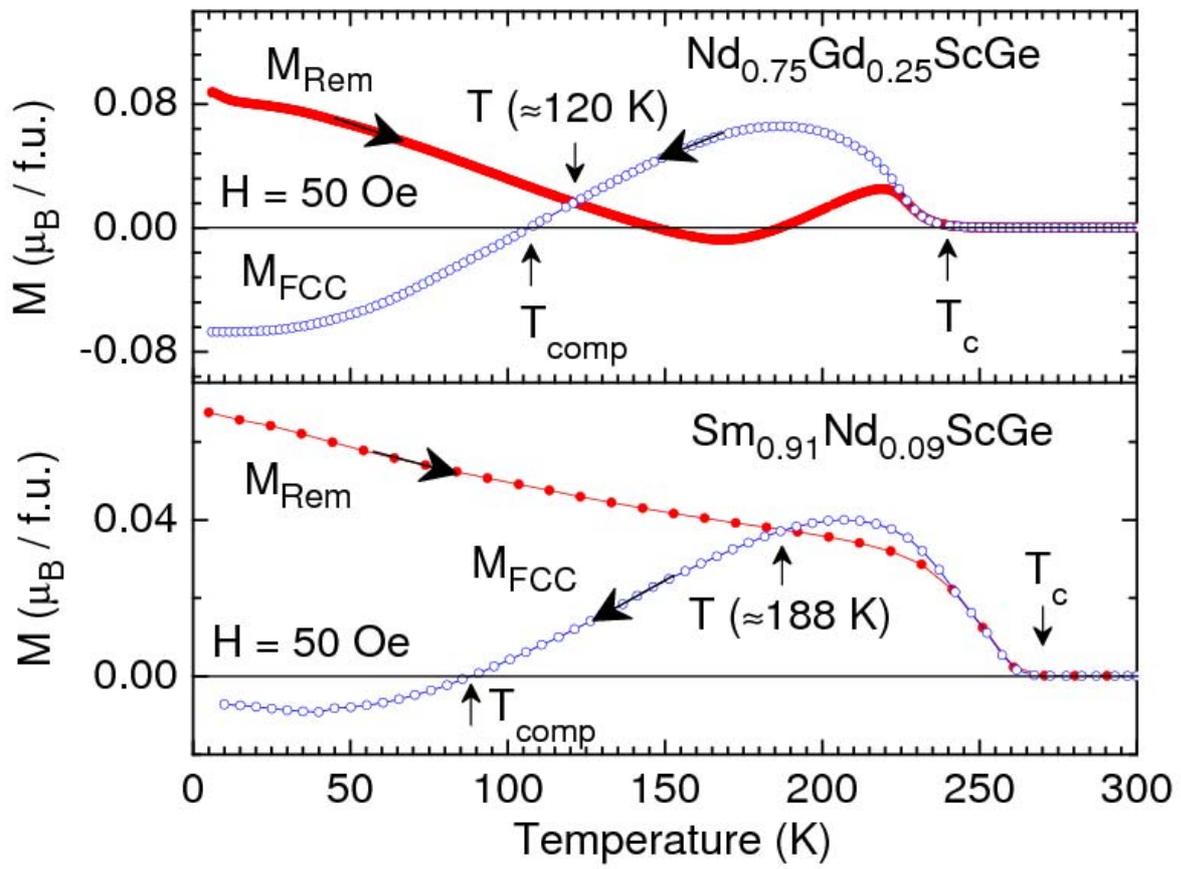

Fig. 3.

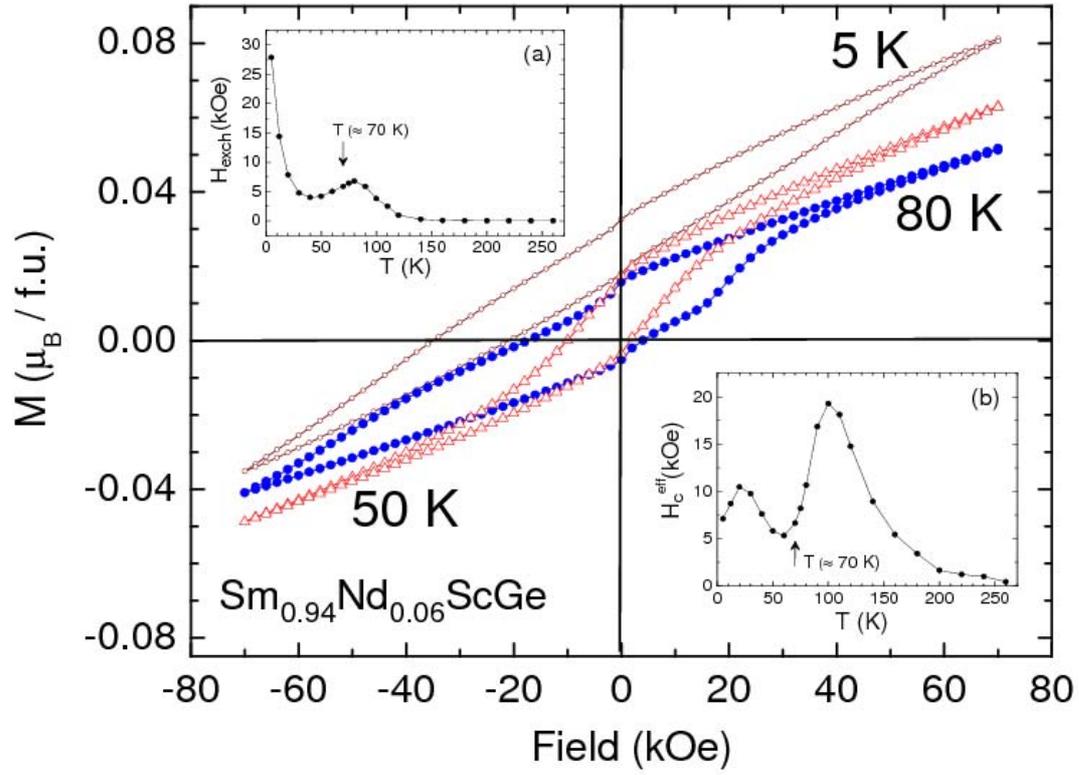

Fig. 4.

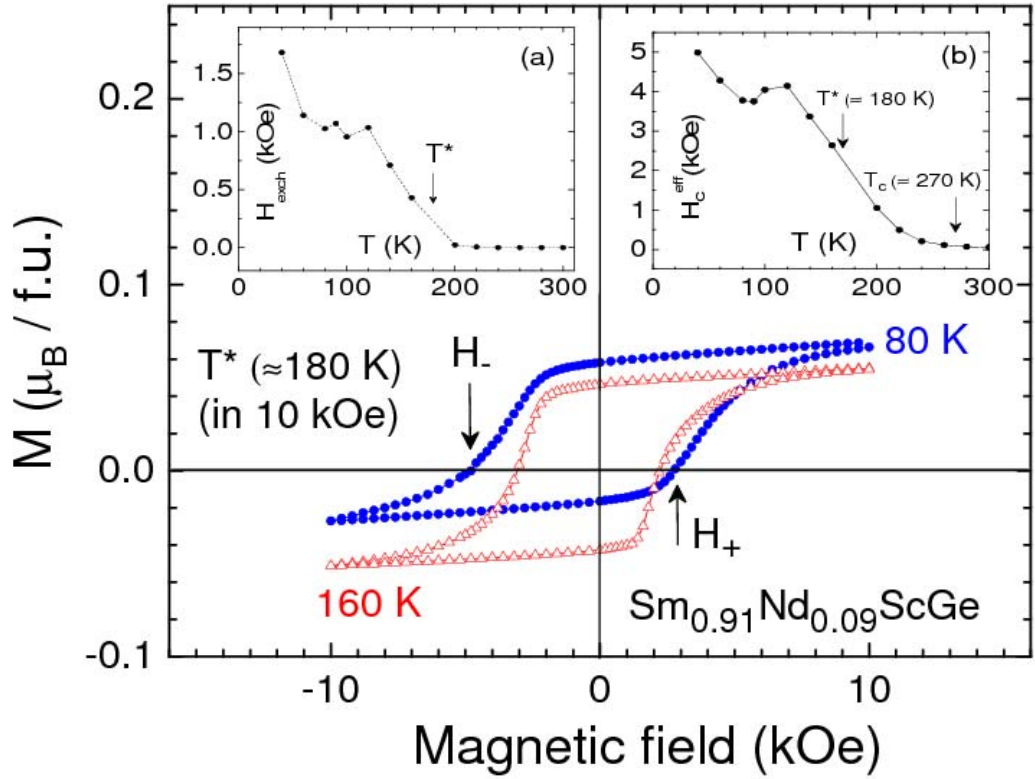

Fig. 5.